# Improving Protein Gamma-Turn Prediction Using Inception Capsule Networks

Chao Fang[1], Yi Shang[1,*] and Dong Xu[1,2,*]

[1]Department of Electrical Engineering and Computer Science, University of Missouri, Columbia, Missouri 65211, USA

[2]Christopher S. Bond Life Sciences Centre, University of Missouri, Columbia, Missouri 65211, USA

*To whom correspondence should be addressed.

**Abstract**
Protein gamma-turn prediction is useful in protein function studies and experimental design. Several methods for gamma-turn prediction have been developed, but the results were unsatisfactory with Matthew correlation coefficients (MCC) around 0.2-0.4. One reason for the low prediction accuracy is the limited capacity of the methods; in particular, the traditional machine-learning methods like SVM may not extract high-level features well to distinguish between turn or non-turn. Hence, it is worthwhile exploring new machine-learning methods for the prediction. A cutting-edge deep neural network, named Capsule Network (CapsuleNet), provides a new opportunity for gamma-turn prediction. Even when the number of input samples is relatively small, the capsules from CapsuleNet are very effective to extract high-level features for classification tasks. Here, we propose a deep inception capsule network for gamma-turn prediction. Its performance on the gamma-turn benchmark GT320 achieved an MCC of 0.45, which significantly outperformed the previous best method with an MCC of 0.38. This is the first gamma-turn prediction method utilizing deep neural networks. Also, to our knowledge, it is the first published bioinformatics application utilizing capsule network, which will provides a useful example for the community.
**Contact:** cf797@mail.missouri.edu, shangy@missouri.edu, xudong@missouri.edu

## 1 Introduction

Protein tertiary structure prediction has been an active research topic since half a century ago (Dill and MacCallum, 2012; Zhou *et al.,* 2011; Webb and Sali, 2014). Because it is challenging to directly predict the protein tertiary structure from a sequence, it has been divided into some sub-problems, such as protein secondary and super-secondary structure predictions. Protein secondary structure consists of three elements such as alpha-helix, beta-sheets and coil (Richardson, 1981). The coils can be classified into tight turns, bulges and random coil structures (Milner-White *et al.,* 1987). Tight turns can be further classified into alpha-, gamma-, delta-, pi- and beta -turns based on the number of amino acids involved in forming the turns and their features (Rose *et al.,* 1985). The tight turns play an important role in forming super-secondary structures and global 3D structure folding.

Gamma-turns are the second most commonly found turns (the first is beta-turns) in proteins. By definition, a gamma-turn contains three consecutive residues (denoted by $i$, $i+1$, $i+2$) and a hydrogen bond between the backbone $CO_i$ and the backbone $NH_{i+2}$ (see Figure 1). There are two types of gamma-turns: classic and inverse (Bystrov *et al.*, 1969). According to (Guruprasad and Rajkumar, 2000), gamma-turns account for 3.4% of total amino acids in proteins. Gamma-turns can be assigned based on protein 3D structures by using Promotif software (Hutchinson and Thornton, 1994). There are two types of gamma-turn prediction problems: (1) gamma-turn/non-gamma-turn prediction (Guruprasad *et al.,* 2003; Kaur and Raghava, 2002; Pham *et al.,* 2005), and (2) gamma-turn type prediction (Chou, 1997; Chou and Blinn 1999; Jahandideh *et al.,* 2007).

The previous methods can be roughly classified into two categories: statistical methods and machine-learning methods. Early predictors (Alkorta *et al.*, 1994; Kaur and Raghava, 2002; Guruprasad *et al*, 2003) built statistical model and machine-learning method to predict gamma-turns. For example, Garnier *et al* (1978), Gibrat *et al* (1987), and Chou (1997) applied statistical models while Pham *et al.* (2005) employed support vector machine (SVM). The gamma-turn prediction has improved gradually, and the improvement comes from both methods and features used. Chou and Blinn (1997) applied a residue-coupled model and achieved prediction MCC 0.08. Kaur and Raghava (2003) used multiple sequence alignments as the feature input and achieved MCC 0.17. Hu and Li (2008) applied SVM and achieved MCC 0.18. Zhu *et al.* (2012) used shape string and position specific scoring matrix (PSSM) from PSIBLAST as inputs and achieved MCC 0.38, which had the best performance prior to this study. The machine-learning methods outperformed statistical methods greatly. However, the gamma-turns prediction performance is still quite low due to several reasons: (1) the dataset is very imbalanced, with about 30:1 of non-gamma-turns and gamma-turns; (2) gamma-turns are relatively rare in proteins, yielding a small training sample size; (3) previous machine-learning methods have not fully exploited the relevant features of gamma-turns. The deep-learning framework may provide a more powerful approach for protein sequence analysis and prediction problems. (Fang *et al.,* 2017; Fang *et al.,* 2018; Fang *et al.,* 2018; Wang *et al.,* 2017) than previous machine-learning techniques.

The recent deep neural networks have achieved outstanding performance in image recognition tasks, using methods such as Inception networks (Szegedy *et al.,* 2017). The main components in the Inception networks are inception blocks, each of which contains stacks of Convolutional Neural Networks (CNNs) (Szegedy *et al.,* 2017). The CNN neurons are scalar, which may not be able to fully capture the spatial relationship between extracted high-level features. To tackle this problem, Sabour *et al.* (2017) proposed a novel deep-learning architecture, named

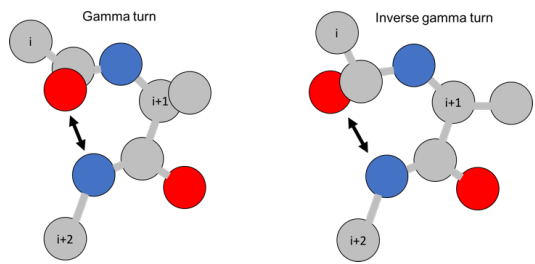

**Fig. 1.** An illustration of gamma-turns. Red circles represent oxygen; grey circles represent carbon; and blue circles represent nitrogen.

Capsule Network (CapsuleNet). The main components of CapsuleNet are capsules, which are groups of neuron vectors. The dimensions of a vector represent the characteristics of patterns, while the length (norm) of a vector represents the probability of existence. In (Sabour *et al.*, 2017), a CapsuleNet was trained for digit classification tasks and the length of a digit capsule represents the confidence of a certain digit being correctly classified and the dimensions of this digit capsule represents different features, such as the stroke thickness, skewness, and scale of a digit image. CapsuleNet, compared to CNN, better captures the structural relationship among the high-level extracted features.

Although CapsuleNets were primarily developed to capture orientation and relative position information of 'entities' or 'objects' in an image, in this paper we apply CapsuleNet to the biological sequence analysis problem from a different perspective. The motivation for applying CapsuleNet for gamma-turn prediction is due to its good properties: First, the dimension of a capsule can be used to reflect certain sequence properties of forming a gamma-turn. The capsule length also gives the confidence or prediction reliability of a predicted gamma-turn label. For example, the closer a capsule length (its norm value) is to 1, the more confident a predicted gamma-turn label is. Second, CapsuleNet contains capsules, each of which can detect a specific type of entity (Sabour *et al.*, 2017). For an MNIST digit recognition task, each capsule was used to detect one class of digits, i.e. the first digit capsule detects 1's; similarly, in this work, each capsule will be used to detect whether it is a classical turn, an inverse turn or non-turn. Also, compared to CNN which has invariance property, CapsuleNet has the equivariance property. The equivariance property means that a translation of input features results in an equivalent translation of outputs, which enables the network to generate features from different perspectives and hence requires a smaller sample size to train than previous CNN architectures. This is useful for many bioinformatics problems: even when the labelled data are scarce and limited, CapsuleNet can detect some high-level features and use them for robust classification. Third, the dynamic routing in CapsuleNet is similar to the attention mechanism (Bahdanau *et al.,* 2014). The routing by agreement mechanism will let a lower-level capsule prefer to send its output to higher-level capsules whose activity vectors have a big scalar product with the prediction coming from the lower-level capsule. In other words, the capsules can "highlight" the most relevant features for a classification task, in this case, gamma-turn classification.

Here, we proposed a deep inception capsule network, which combines CapsuleNet with inception network for protein gamma-turn prediction. First, we performed extensive experiments to test the CapsuleNet performance under different conditions. Next, we show that the proposed network outperformed previous predictors utilizing traditional machine-learning methods such as SVM on public benchmarks. Last but not least, we further explored the features learnt by capsules and connected them back to the protein sequence to discover useful motifs that may form a gamma turn.

## 2 Materials and Methods

### 2.1 Problem formulation

A protein gamma-turn prediction is a binary classification problem, which can be formulated as followed: given a primary sequence of a protein, a sliding window of $k$ residues were used to predict the central residue turn or non-turn. For example, if $k$ is 17, then each protein is subsequently sliced into fragments of 17 amino acids with a sliding window.

To make accurate prediction, it is important to provide useful input features to machine-learning methods. We carefully designed a feature matrix corresponding to the primary amino acid sequence of a protein, which consists of a rich set of information derived from individual amino acid, as well as the context of the protein sequence. Specifically, the feature matrix is a composition of HHBlits profile (Remmert *et al.,* 2012), and predicted protein shape string using Frag1D (Zhou *et al.,* 2009).

The first set of useful features comes from the protein profiles generated using HHBlits (Remmert *et al.*, 2012). In our experiments, the HHBlits software used database uniprot20_2013_03, which can be downloaded from http://wwwuser.gwdg.de/~compbiol/data/hhsuite/databases/hhsuite_dbs/. A HHBlits profile can reflect the evolutionary information of the protein sequence based on a search of the given protein sequence against a sequence database. The profile values were scaled by the sigmoid function into the range (0, 1). Each amino acid in the protein sequence is represented as a vector of 31 real numbers, of which 30 from HHM Profile values and 1 *NoSeq* label (representing a gap) in the last column. The HHBlits profile corresponds to amino acids and some transition probabilities, i.e., A, C, D, E, F, G, H, I, K, L, M, N, P, Q, R, S, T, V, W, Y, M->M, M->I, M->D, I->M, I->I, D->M, D->D, Neff, Neff_I, and Neff_D.

The second set of useful features, predicted shape string, comes from Frag1D (Zhou *et al.,* 2009). For each protein sequence, Frag1D can generate useful predicted protein 1D structure features: classical three-state secondary structures, and three- and eight-state shape strings. Classical three-state secondary structures and three-state shape string labels both contain H (helix), S (sheet), and R (random loop), but they are based on different methods so that they have small differences. In this experiment, we used all the features from Frag1D. Eight-state shape string labels contain R (polyproline type alpha structure), S (beta sheet), U/V (bridging regions), A (alpha helices), K ($3_{10}$ helices), G (almost entirely glycine), and T (turns). The classical prediction of three-state protein secondary structures has been used as an important feature for protein structure prediction, but it does not carry further structural information for the loop regions, which account for an average of 40% of all residues in proteins. Ison *et al.* (2005) proposed Shape Strings, which give a 1D string of symbols representing the distribution of protein backbone psi-phi torsion angles. The shape strings include the conformations of residues in regular secondary structure elements; in particular, shape 'A' corresponds to alpha helix and shape 'S' corresponds to beta strand. Besides, shape strings classify the random loop regions into several states that contain much more conformational information, which we found particularly useful for gamma-turn prediction problem. For the Frag1D prediction result, each amino acid in the protein sequence is represented as a vector of 15 numbers, of which 3 from the classical three-state secondary structures, 3 from the three-state shape strings, 8 from the eight-state shape strings and 1 *NoSeq* label in the last column. The predicted classical three-state secondary structure feature is represented as one-hot encoding as followed: helix: (1,0,0), strand: (0,1,0), and loop: (0,0,1). The same rule applies to three- and eight-state shape string features. In this work, we also tried the traditional eight-state protein secondary structures. However, the prediction result was not as good as the one from the eight-state shape strings. This is probably because the traditional eight-state secondary structures contain much less structural information for the gamma-turn prediction problem.

## 2.2 Model design

In this section, a new deep inception capsule network (DeepICN) is presented. Figure 2A shows the model design. The input features for DeepICN are HHBlits profiles and predicted shape strings. Since the distributions of HHBlits profiles and predicted shape strings are different, we applied convolutional filters separately on the two features, then concatenated them. The CNN is used to generate the convolved features. We first applied CNN to extract local low-level features from protein profiles and predicted shape strings features. This CNN layer will extract local features similar to a CNN used to extract "edge" features of objects in an image (Xie and Tu, 2015).

After the convoluted feature concatenation, the merged features are fed into the inception module (See Figure 2B for details). The inception network (Szegedy *et al.*, 2017) was then applied to extract low-to-intermediate features for CapsuleNet. In (Sabour *et al.*, 2017), CapsuleNet was used for digital image classification and the primary capsule layers were placed after a convolutional layer. Their network design worked well for digital image recognition with the image dimension 28-by-28. Considering the complex features of protein HHblits profile and shape strings, it is reasonable to apply a deeper network to extract local to medium level features so that CapsuleNet can work well on top of those features and extract high-level features for gamma-turn classification. The purpose of setting up an inception block right after CNN is to extract intermediate-level features.

Each convolution layer, such as 'Conv (3)' in Figure 2B, consists of four operations in sequential order: (1) a one-dimensional convolution operation using the kernel size of three; (2) the batch normalization technique (Ioffe and Szegedy, 2015) for speeding up the training process and acting as a regularizer; (3) the activation operation, ReLU (Radford *et al.*, 2015); and (4) the dropout operation (Srivastava *et al.*, 2014) to prevent the neural network from overfitting by randomly dropping neurons during the deep network training process so that the network can avoid co-adapting.

The capsule layers are placed after the inception module to extract high-level features or explore the spatial relationship among the local features that are extracted in the above-mentioned layers. The primary capsule layer (See Figure 2C) is a convolutional capsule layer as described in (Sabour *et al.*, 2017). It contains 32 channels of convolutional 8D capsules, with a 9 x 9 kernel and a stride of 2. The final layer (turn capsule) has two 16D capsules to represent two states of the predicted label: gamma-turn or non-gamma-turn. The weights between primary capsules and turn capsules are determined by the iterative dynamic routing algorithm (Sabour *et al.*, 2017). The squashing activation function (Sabour *et al.*, 2017) was applied in the computation between the primary capsule layer and the turn capsule layer as follows:

$$v_j = \frac{\|s_j\|}{1 + \|s_j\|^2} \frac{s_j}{\|s_j\|}$$

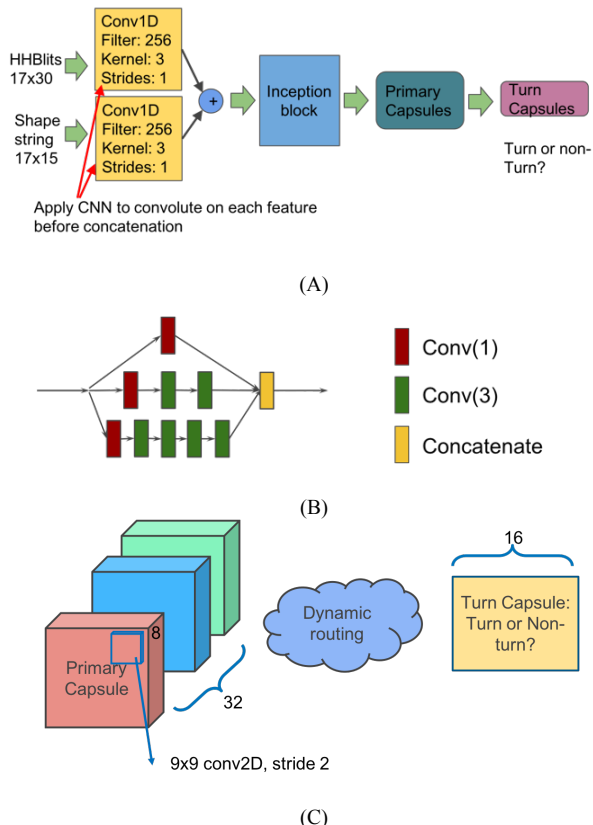

(A)

(B)

(C)

**Fig. 2.** (A) A deep inception capsule network design. The input features are HHBlits profile (17-by-30 2D array) and predicted shape string using Frag1D (17-by-15 2D array). Each feature is convolved by a convolutional layer. Both convolved features then get concatenated. An inception block is followed to extract low to medium features. A primary capsule layer then extracts higher level features. The final turn capsule layer makes predictions. (B) An inception block. Inside this inception block: Red square Conv(1) stands for convolution operation with kernel size 1. Green square Conv(3) stands for convolution operation with kernel size 3. Yellow square stands for feature map concatenation. (C) Zoom-in between primary capsules and turn capsules. The primary capsule layer contains 32 channels of convolutional 8D capsules. The final layer turn capsule has two 16D capsules to represent two states of the predicted label: gamma-turn or non-gamma-turn. The computation between those two layers is dynamic routing.

where $v_j$ is the vector output of capsule $j$ and $s_j$ is its total output. The dynamic routing algorithm (Sabour *et al.*, 2017) is as follows:

Routing Algorithm:

Routing $(\hat{u}_{j|i}, r, l)$

    For all capsule $i$ in layer $l$ and capsule $j$ in layer $(l+1)$: $b_{i,j} \leftarrow 0$

    For $r$ iteration do

        For all capsule $i$ in layer $l$: $c_i \leftarrow softmax(b_i)$

        For all capsule $j$ in layer $(l+1)$: $s_j \leftarrow \sum_i c_{ij} \hat{u}_{j|i}$

        For all capsule $j$ in layer $(l+1)$: $v_j \leftarrow squash(s_j)$

        For all capsule $i$ in layer $l$ and capsule $j$ in layer $(l+1)$: $b_{ij} \leftarrow b_{ij} + \hat{u}_{j|i} \cdot v_j$

The evaluation matric for gamma-turn prediction more commonly uses Matthew Correlation Coefficient (MCC) than percentage accuracy since the accuracy only considers the true positives and false positives without the true negatives and false negatives. Another reason is that the gamma-turn dataset is very imbalanced, where MCC can evaluate how well the classifier performs on both positive and negative labels. MCC can be calculated from the confusion matrix as follows:

$$MCC = \frac{TP*TN - FP*FN}{\sqrt{(TP+FP)(TP+FN)(TN+FP)(TN+FN)}}$$

where TP is the number of true positives, TN is the number of true negatives, FP is the number of false positives and FN is the number of false negatives.

### 2.3 Model training

DeepICN was implemented, trained, and tested using TensorFlow and Keras. Different sets of hyper-parameters (dynamic routing iteration times, training data sample size, convolution kernel size, and sliding window size) of DeepICN were explored. An early stopping strategy was used when training the models and if the validation loss did not reduce in 10 epochs, the training process stopped. The Adam optimizer was used to dynamically change the learning rate during model training. All the experiments were performed on an Alienware Area-51 desktop equipped with a Nvidia Titan X GPU (11 GB graphic memory).

### 2.4 Experiment Dataset

**1) CullPDB** (Wang and Dunbrack, 2003) was download on November 2, 2017. It originally contained 20,346 proteins with percentage cutoff 90% in sequence identity, resolution cutoff 2.0 Å, and R-factor cutoff 0.25. This dataset was preprocessed and cleaned up by satisfying all the following conditions: with length less than 700 amino acids; with valid PSIBLAST profile, HHblits profile; with shape strings predicted by Frag1D (Zhou *et al.*, 2009); and with gamma-turn labels retrieved by PROMOTIF (Hutchinson and Thornton, 1996). After this, 19,561 proteins remained and CD-Hit (Li and Godzik, 2006) with 30% sequence identity cutoff was applied on this dataset resulting in 10,007 proteins. We removed proteins with sequence identity more than 30% for an objective and strict test in terms of model generalization. This dataset was mainly used for deep neural network hyper-parameter tuning and the exploration of CapsuleNet configurations. It was also used to compare the proposed inception capsule model with other deep-learning models. For these purposes, a balanced dataset was built: all positive gamma-turn labels and an equal size of negative non-gamma-turn labels were selected to form a balanced dataset.

**2)** The benchmark **GT320** (Guruprasad and Rajkumar, 2000) is a common data set used for benchmarking gamma-turn prediction methods. GT320 contains 320 non-homologous protein chains in total with 25% sequence identity cutoffs, and resolution better than 2.0 Å resolution. This benchmark was used to compare the performance with previous predictors. Each chain contains at least one gamma-turn. The gamma-turns were assigned by PROMOTIF (Hutchinson and Thornton, 1996). In this work, five-fold cross-validation experiments on GT320 was performed and results were compared against other predictors.

## 3 Experimental Results

In this section, extensive experimental results of the proposed deep CapsuleNets with different hyper-parameters were tuned and tested using CullPDB and five-fold cross-validation results on GT320. The performance comparison with existing methods is presented.

**Table 1.** Effect of window size on MCC performance

| Window size | Test average MCC | Time (hr) | P-value on MCC |
|---|---|---|---|
| 15 | 0.4458(±0.0107) | 0.18(±0.11) | 0.0115 |
| 17 | 0.4645(±0.0062) | 0.24(±0.15) | - |
| 19 | 0.4442(±0.0049) | 0.37(±0.18) | 0.0010 |
| 21 | 0.4548(±0.0055) | 0.43(±0.20) | 0.0499 |
| 23 | 0.4227(±0.0076) | 0.37(±0.23) | 0.0001 |
| 25 | 0.4369(±0.0076) | 0.45(±0.25) | 0.0005 |

### 3.1 Hyper-parameter Tuning and Model Performance

Tables 1-4 show the exploration of the inception capsule network with different hyper-parameters. This set of experiments was to find out a better configuration of hyper-parameters for the deep networks using the CullPDB dataset. Since this network involves many hyper-parameters, only the major ones were explored. Table 1 shows how the sliding window size affects the model performance. In this experiment, 1000 proteins were randomly selected to form the training set, 500 for the validation set and 500 for the test set. Each experiment was performed with five times of data randomization.

Table 1 shows how the sliding window size of input affects the deep capsule network performance. The larger the window size, the more training time it took for CapsuleNet. However, MCC may not grow as the window size increases. We chose the window size of 17 amino acids based on its peak MCC performance in the experiments. The t-test p-values show that the window size 17 test MCC compared to other window sizes is statistically significant.

Table 2 shows the dropout rate's effects on the performance of the deep capsule network. If a dropout was not used, the network had very high over-fitting. The dropout rate 0.4-0.5 is reasonable as it is a compromise between the training and test prediction performance. We chose dropout 0.5 in our study. The P-value between the dropout of 0.5 and any of others was insignificant. Although The dropout of 0.8 had the highest test average MCC, its standard deviation (±0.0249) is also high, and hence, we did not use it.

**Table 2.** Effect of dropout on MCC performance

| Dropout | Train average MCC | Test average MCC | P-value on test MCC |
|---|---|---|---|
| No | 0.9974(±0.0015) | 0.4439(±0.0101) | 0.1236 |
| 0.3 | 0.9857(±0.0154) | 0.4454(±0.0049) | 0.0843 |
| 0.4 | 0.9010(±0.1457) | 0.4515(±0.0047) | 0.4294 |
| 0.5 | 0.9377(±0.0598) | 0.4558(±0.0092) | - |
| 0.6 | 0.9159(±0.0688) | 0.4525(±0.0111) | 0.6647 |
| 0.7 | 0.8371(±0.0920) | 0.4604(±0.0063) | 0.4318 |
| 0.8 | 0.6072(±0.1033) | 0.4646(±0.0249) | 0.5228 |

**Table 3.** Effect of training size on training time and MCC performance

| Training size | Test average MCC | Time (hr) |
|---|---|---|
| 500 | 0.4224(±0.0035) | 0.23(±0.17) |
| 1000 | 0.4553(±0.0098) | 0.87(±0.03) |
| 2000 | 0.4422(±0.0204) | 1.59(±0.07) |
| 3000 | 0.4752(±0.0111) | 2.38(±0.09) |
| 4000 | 0.4787(±0.0147) | 3.13(±0.12) |
| 5000 | 0.4717(±0.0165) | 3.91(±0.14) |

**Table 4.** Effect of dynamic routing on MCC performance

| dynamic routing times | Test average MCC | Time (hr) | P-value on MCC |
|---|---|---|---|
| 1 | 0.4454(±0.0049) | 0.44(±0.16) | 0.4644 |
| 2 | 0.4492(±0.0086) | 0.31(±0.17) | - |
| 3 | 0.4407(±0.0032) | 0.37(±0.15) | 0.1017 |
| 4 | 0.4497(±0.0045) | 0.32(±0.18) | 0.9276 |
| 5 | 0.4487(±0.0061) | 0.41(±0.14) | 0.9502 |

Table 3 shows the effects of the training sample size on the deep capsule network training speed and performance. More training data increased and training time and the model performance. However, after 3000 samples, the MCC performance did not improve significantly with more training data. This is consistent with the observation in (Sabour *et al.*, 2017) that CapsuleNet did not need a large dataset for training.

Table 4 shows the effect of number of dynamic routing on the performance. Dynamic routing is used in CapsuleNet similar to max-pooling in a CNN, but it is more effective than max-pooling in that it allows neurons in one layer to ignore all but the most active feature detector in a local pool in the previous layer. In this experiment, we fixed the other hyper-parameters searched in the above-mentioned experiments and studied how number of dynamic routing affected the performance. Considering the training time and the MCC performance, 2 routings are suitable, as more dynamic routing does not have significant improvement. The training time did not show large variations as the number of dynamic routings increases. This may be because our experiments used early stopping.

### 3.2 Prediction confidence: the capsule length

According to (Sabour *et al.*, 2017), the capsule length indicates the probability that the entity represented by the capsule is present in the current input. In other words, the capsule length in the last layer can be used for prediction of gamma-turn and assessment of prediction confidence. The longer the turn capsule length is, the more confident the prediction of a turn capsule will be. Here, the capsule length in Turn Capsules can be used to show how confidence a gamma-turn is predicted. Specifically, a test set (with 5000 proteins containing 19,594 data samples) was fed into the trained inception capsule network to get a capsule length vector. Then the capsule length vector that represents positive capsules were kept. Since all the capsule length values fall into the range between 0 and 1, they were grouped into bins with the width of 0.05, so that there are totally 20 bins. The precision of each bin can be calculated to represent the prediction confidence. Figure 3 shows the fitting curve of precision (percentage of correctly predicted gamma-turns, i.e., true positives in the bin) versus the capsule length. A nonlinear regression curve was used to fit all the points, yielding the following equation:

$$y = 1.084x^2 - 0.203x + 0.147$$

where $x$ is the capsule length and $y$ is the precision.

The fitting-curve can be further used for prediction confidence assessment: given a capsule length, its prediction confidence can be estimated using the above equation.

### 3.3 Proposed model performance compared with previous predictors

For comparing with other predictors, the public benchmark GT320 was used. Following the previous studies, a five-fold cross validation results were reported, as shown in Table 5. This GT320 is an imbalanced dataset, but for objective evaluation, we did not sample any balanced data from training or testing, as done in previous studies.

**Table 5.** Performance comparison with previous predictors using GT320 benchmark.

| Methods | MCC |
|---|---|
| Our Approach | 0.45 |
| Zhu *et al.*, 2012 | 0.38 |
| Hu's SVM | 0.18 |
| SNNS | 0.17 |
| GTSVM | 0.12 |
| WEKA-logistic regression | 0.12 |
| WEKA-naïve Bayes | 0.11 |

*The results of WEKA, SNNS were obtained from (Kaur and Raghava, 2002), the result of GTSVM was obtained from (Pham *et al.*, 2005) and result of Hu's SVM was from (Hu and Li, 2008). Zhu *et al.*, (2012) is the previous best predictor result.

Table 5 shows that the proposed inception capsule network outperformed all the previous methods by a significant margin.

### 3.4 Extend CapsuleNet for classic and inverse gamma-turn prediction

Many previous gamma-turn predictors only predict whether a turn is gamma-turn or not. Here, we also extended our deep inception capsule model for classic and inverse gamma-turn prediction. The experiment dataset is still CullPDB, and inverse and classic labels were assigned using

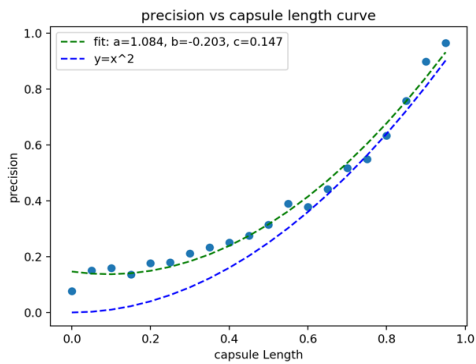

**Fig. 3.** The fitting curve of precision (percentage of true positive in the bin) versus the capsule length. The green line is the fitting curve and the blue line (y=x^2) is for reference.

PROMOTIF (Hutchinson and Thornton, 1996). The same deep inception capsule network (shown in Figure 2A) was applied except the last turn capsule layer now has three capsules to predict non-turn, inverse turn or classic turn as a three-class classification problem. The performance metric Q3 is used which is the accuracy of correct prediction for each class. The prediction results are shown in Table 6. Different numbers of proteins were used to build the training set. The validation and test set contain 500 proteins each. The CullPDB dataset contains 10,007 proteins which have 1383 classic turns, 17,800 inverse turns, and 2,439,018 non-turns in total. This is a very imbalanced dataset. In this experiment, the balanced training set, validation set, and test set were generated as follows: The inverse turn samples were randomly drawn as much as classic turn sample size. For the non-turn samples, they were randomly drawn twice as much as classic turn sample size, i.e. the sum of inverse turn samples and classic turn samples. The training loss and validation loss curves are shown in Figure 4. From the loss curve, it shows that after about 75 epochs, the model learning process was converging. Since the model hyper-parameters had been explored in the earlier experiments, during this experiment, we adopted similar values, i.e., the window size was chosen 17 amino acids, the filter size is 256, the convolution kernel size was chosen 3, the dynamic routing was chosen 3 iterations and dropout ratio was 0.3.

**Table 6.** Non-turn, inverse and classic turn prediction results.

| Training size | Test average Q3 | Time (hr) | P-value |
|---|---|---|---|
| 5000 | 0.6839(±0.0053) | 0.25(±0.20) | - |
| 6000 | 0.6783(±0.0076) | 0.38(±0.22) | 0.2706 |
| 7000 | 0.6864(±0.0124) | 0.34(±0.16) | 0.3057 |

### 3.5 Visualization of the features learnt by capsules

In order to verify whether the extract high-level features learnt/extracted from the input data have the prediction power and are generalizable, t-SNE (Maaten and Hinton, 2008) was applied to visualize the input features and the capsule features for both the training data and the test data. Figure 5 (A) shows the t-SNE plot of the input features from the training data before the training. The input data has 45 features (i.e. 45 dimensions), and t-SNE can project 45 dimensions onto two principal dimensions and visualize it. There was no clear cluster in the training data. Figure 5 (B) shows the t-SNE plot of the capsule features from the training data. The turn capsule contains 16 dimensions, and the t-SNE can similarly project the capsule features to two major principal features and visualize it. The clusters were obviously formed after the training. Figure 5 (C) and (D) show the t-SNE plots for the input features and the capsule features of the test data. There was no clear cluster for the input features in the test data either. The capsule features still tends to be clustered together in the test data, although to less extent than the training data.

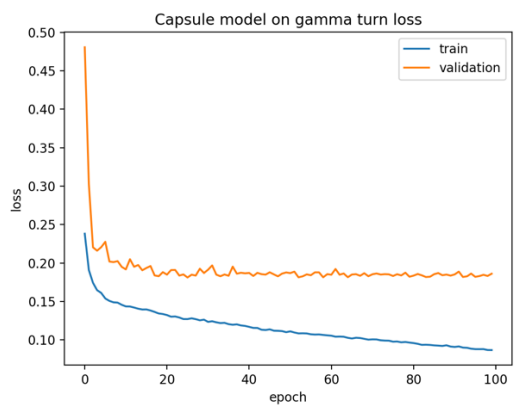

**Fig. 4.** Training loss and validation loss curve of deep inception capsule network for classic and inverse gamma-turn.

Figure 6 (A) shows the classic turn Weblogo and Figure 6 (B) shows the inverse turn Weblogo (Crooks *et al.*, 2004). In the two plots, the y axis has the same height of 0.8 bits. Both types of turns have some visible features and the classic turn Weblogo contains more information content than the inverse turn.

### 3.6 Ablation study

To discovery the important elements in our proposed network, an ablation study was performed by removing or replacing different components in the deep inception capsule network. In particular, we tested the performance of the proposed models without the capsule component, replacing the capsule component with CNN, or replacing inception component with CNN. Each ablation experiment was performed using the same allocation of the data (3000 proteins for training, 500 proteins for validation, and 500 for test) and same parameter setting: dropout ratio 0.5 and window size 17. From the ablation test result presented in Table 7, we found that the capsule component is the most effective component in our network, since the performance dropped significantly when removing or replacing the capsule component. The inception component also acts as an important component as it can effectively extract feature maps for capsule components compared to CNN.

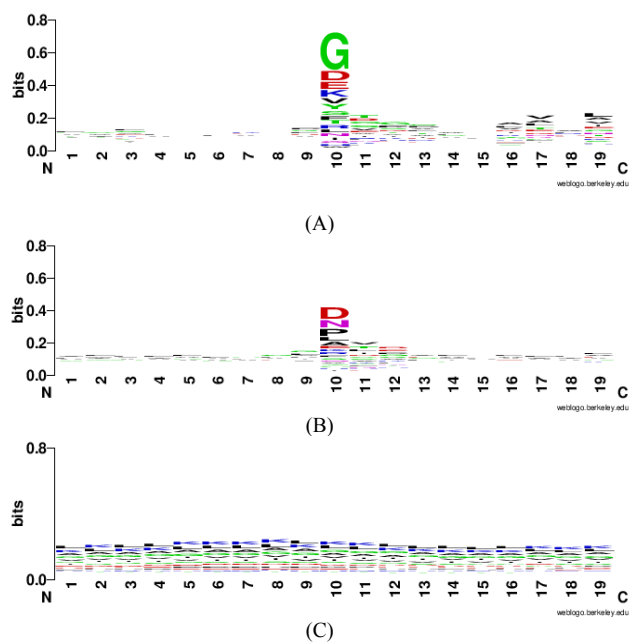

**Fig. 6.** (A) Classic turn Weblogo; (B) inverse turn Weblogo; and (C) Non-turn Weblogo.

**Table 7.** Ablation test.

| Model | MCC |
| --- | --- |
| Replace inception component with CNN | 0.4544(±0.0106) |
| Replace capsule component with CNN | 0.4485(±0.0056) |
| Without capsule component | 0.4551(±0.0059) |
| Proposed Design | 0.4752(±0.0111) |

## 4 Conclusion and Discussion

In this work, the latest deep-learning framework, CapsuleNet, was applied to protein gamma-turn prediction. Instead of applying capsule network directly, a new model called inception capsule network was proposed and has shown improved performance comparing to previous predictors. This work has several innovations.

First of all, this work is the first application of deep neural networks to protein gamma-turn prediction. Compared to previous traditional machine-learning methods for protein gamma-turn prediction, this work uses a more sophisticated, yet efficient, deep-learning architecture, which outperforms previous methods. A software tool has been developed and it will provide the research community a powerful deep-learning prediction tool for gamma-turn prediction. The ablation test was performed, and the importance of capsule component was verified.

Second, this work is the earliest application of CapsuleNet to protein structure-related prediction, and possibly to any bioinformatics problems to our knowledge, as CapsuleNet was just published in 2017. Here, we proposed an inception capsule network for protein gamma-turn prediction and explored some unique characters of capsules. To explore the capsule length, we designed an experiment of grouping each capsule length into several bins and discovered the relationship between prediction precision and capsule length. A nonlinear curve can be applied to fit the data and further used for estimating the prediction confidence. In addition, the network was extended to inverse turn and classical turn prediction. The inverse turn capsule and classical turn capsule were further explored by showing the t-SNE visualization of the learnt capsule features. Some interesting motifs were visualized by Weblogo.

Third, new features have been explored and applied to gamma-turn prediction. The features used for network training, namely HHBlits profiles and predicted shape string, contain high information content making deep learning very effective. The HHBlits profiles provide evolutionary information while shape strings provide complementary structural information for effectively predicting gamma turns.

Last but not least, previous gamma-turn resources are very limited and outdated. A few servers are not maintained, and no downloadable executable of gamma-turn is available. Here a free tool with source code utilizing deep learning and state-of-the-art CapsuleNet will be ready for researchers to use.

## Acknowledgments

This work was partially supported by National Institutes of Health grant R01-GM100701. We like to thank Drs. Janet Thornton and Roman Laskowski for their help on compiling and configuring the PROMOTIF software. We also like to thank Duolin Wang, Shuai Zeng, and Zhaoyu Li for their helpful discussions on Capsule Networks.

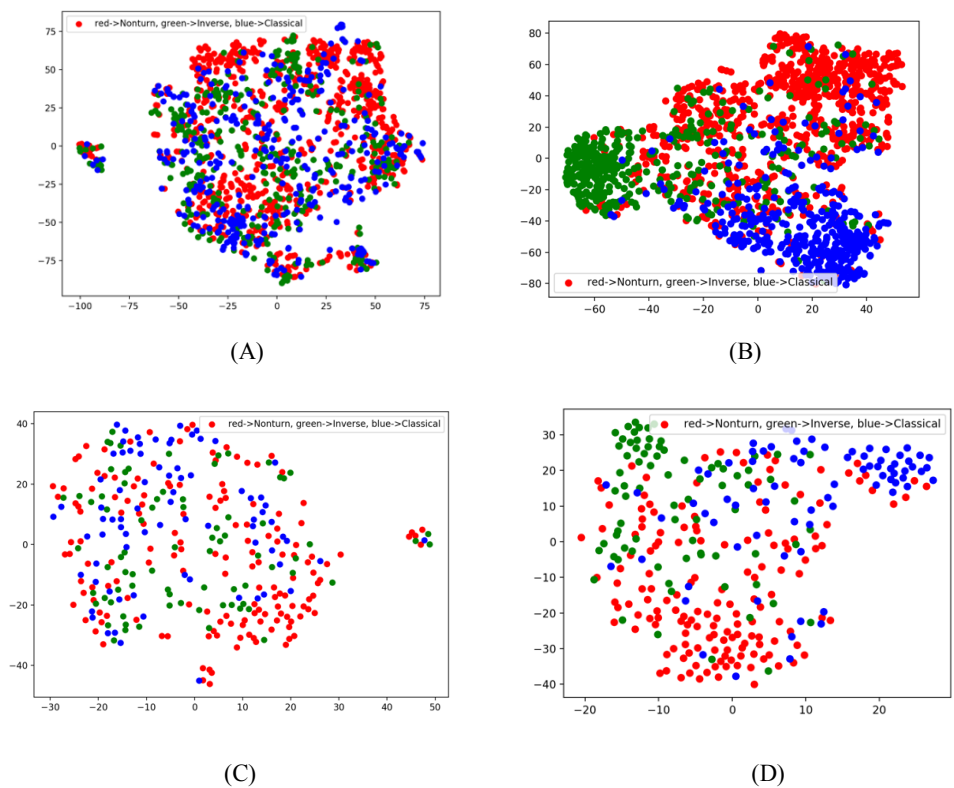

(A)  (B)

(C)  (D)

**Fig. 5.** t-SNE plots of capsule network features. **(A)** and **(B)** are plots of the input features and the capsule features, respectively for training dataset (3000 proteins with 1516 turn samples). **(C)** and **(D)** are plots of the input features and the capsule features, respectively for the test dataset (500 proteins with 312 turn samples). Red dots represent non-turns, green dots represent inverse turns and blue dots represent classic turns.